\documentclass[useAMS,usenatbib]{mn2e}
\usepackage{natbib}
\usepackage{amsmath,amsfonts}
\usepackage{epsfig}
\usepackage{mathptmx}
\def\reff@jnl#1{{\rm#1\/}}

\def\aj{\reff@jnl{AJ}}                  % Astronomical Journal
\def\araa{\reff@jnl{ARA\&A}}            % Annual Review of Astron and Astrophys
\def\apj{\reff@jnl{ApJ}}                        % Astrophysical Journal
\def\apjl{\reff@jnl{ApJ}}               % Astrophysical Journal, Letters
\def\apjs{\reff@jnl{ApJS}}              % Astrophysical Journal, Supplement
\def\ao{\reff@jnl{Appl.Optics}}         % Applied Optics
\def\apss{\reff@jnl{Ap\&SS}}            % Astrophysics and Space Science
\def\aap{\reff@jnl{A\&A}}               % Astronomy and Astrophysics
\def\aapr{\reff@jnl{A\&A~Rev.}}         % Astronomy and Astrophysics Reviews
\def\aaps{\reff@jnl{A\&AS}}             % Astronomy and Astrophysics, Supplement
\def\azh{\reff@jnl{AZh}}                        % Astronomicheskii Zhurnal
\def\baas{\reff@jnl{BAAS}}              % Bulletin of the AAS
\def\jrasc{\reff@jnl{JRASC}}            % Journal of the RAS of Canada
\def\memras{\reff@jnl{MmRAS}}           % Memoirs of the RAS
\def\mnras{\reff@jnl{MNRAS}}            % Monthly Notices of the RAS
\def\pra{\reff@jnl{Phys. Rev. A}}         % Physical Review A: General Physics
\def\prb{\reff@jnl{Phys. Rev. B}}         % Physical Review B: Solid State
\def\prc{\reff@jnl{Phys. Rev. C}}         % Physical Review C
\def\prd{\reff@jnl{Phys. Rev. D}}         % Physical Review D
\def\prl{\reff@jnl{Phys. Rev. Lett}}      % Physical Review Letter
\def\pasp{\reff@jnl{PASP}}              % Publications of the ASP
\def\pasj{\reff@jnl{PASJ}}              % Publications of the ASJ
\def\qjras{\reff@jnl{QJRAS}}            % Quarterly Journal of the RAS
\def\skytel{\reff@jnl{S\&T}}            % Sky and Telescope
\def\solphys{\reff@jnl{Solar~Phys.}}    % Solar Physics
\def\sovast{\reff@jnl{Soviet~Ast.}}     % Soviet Astronomy
\def\ssr{\reff@jnl{Space~Sci.Rev.}}     % Space Science Reviews
\def\zap{\reff@jnl{ZAp}}                        % Zeitschrift fuer Astrophysik
\def\nat{\reff@jnl{Nature}}             % Nature

\def\p#1by#2{{\partial{#1} \over \partial{#2}}}
\def\pp#1by#2#3{{\partial^2{#1} \over \partial{#2}\partial{#3}}}
\def\d#1by#2{{{\rm d}{#1} \over {\rm d}{#2}}}
\def\dd#1by#2#3{{{\rm d}^2{#1} \over {\rm d}{#2}{\rm d}{#3}}}

\title[AMI LA spinning dust observations]{High resolution AMI Large Array
imaging of spinning dust sources: spatially correlated 8\,$\mu$m emission and
evidence of a stellar wind in L675\thanks{We request that any reference to this
paper cites ``AMI Consortium: Scaife et~al.\ 2010''}}

\label{firstpage}

\author[Scaife et~al.]{AMI Consortium:
 Anna M.\ M.\ Scaife$^{1,2}$\thanks{E-mail: ascaife@cp.dias.ie},
 David A.\ Green$^1$, Guy G.\ Pooley$^1$,\newauthor
 Matthew L.\ Davies$^1$, Thomas M.\ O.\ Franzen$^1$,
 Keith J.\ B.\ Grainge$^{1,3}$,\newauthor
 Michael P.\ Hobson$^1$, Natasha Hurley-Walker$^1$,
 Anthony N.\ Lasenby$^{1,3}$, \newauthor
 Malak Olamaie$^1$, John S.\ Richer$^{1,3}$,
 Carmen Rodr{\'i}guez-Gonz{\'a}lvez$^1$, \newauthor
 Richard D.\ E.\ Saunders$^{1,3}$, Paul F.\ Scott$^1$,
 Timothy W.\ Shimwell$^1$, \newauthor
 David J.\ Titterington$^1$, Elizabeth M.\ Waldram$^1$
 \& Jonathan T.\ L.\ Zwart$^4$\\
$^1$ Astrophysics Group, Cavendish Laboratory, J J Thomson Avenue,
     Cambridge CB3 0HE\\
$^2$ Dublin Institute for Advanced Studies, 31 Fitzwilliam Place,
     Dublin 2, Ireland\\
$^3$ Kavli Institute for Cosmology Cambridge, Madingley Road,
     Cambridge, CB3 0HA\\
$^4$ Columbia Astrophysics Laboratory, Columbia University, 550 West 120th
     Street, New York 10027, USA}

\date{Accepted ---; received ---; in original form \today}

\pagerange{\pageref{firstpage}--\pageref{lastpage}}

\pubyear{2009}

\begin{document}
%------------------------------------------------------------------------------%
\maketitle

\begin{abstract}
We present $25''$ resolution radio images of five Lynds Dark Nebulae
(L675, L944, L1103, L1111 and L1246) at 16\,GHz made with the Arcminute
Microkelvin Imager (AMI) Large Array. These objects were previously observed
with the AMI Small Array to have an excess of emission at microwave frequencies
relative to lower frequency radio data. In L675 we find a flat spectrum compact
radio counterpart to the 850\,$\mu$m emission seen with SCUBA and suggest that
it is cm-wave emission from a previously unknown deeply embedded young
protostar. In the case of L1246 the cm-wave emission is spatially correlated
with 8\,$\mu$m emission seen with \emph{Spitzer}. Since the MIR emission is
present only in \emph{Spitzer} band 4 we suggest that it arises from a
population of PAH molecules, which also give rise to the cm-wave emission
through spinning dust emission.
\end{abstract}


\section{Introduction}

The complete characterization of microwave emission from spinning dust grains
is a key question in both astrophysics and cosmology. It probes a region of the
electromagnetic spectrum where a number of different astrophysical disciplines
overlap. It is important for CMB observations in order to correctly
characterise the contaminating foreground emission; for star and planetary
formation it is important because it potentially probes a regime of grain sizes
that is not otherwise easily observable.

Although a number of objects have now been found to exhibit anomalous microwave
emission, attributed to spinning dust, it is still unclear what differentiates
those objects from the many other seemingly similar targets that do not show
the excess. In the specific case of dark clouds the recent AMI sample (AMI
Consortium: Scaife et~al.\ 2009; hereinafter Paper I) of fourteen Lynds Dark
Nebulae found an excess in only five.

It has been suggested that cm-wave emission from spinning dust is emitted by a
population of ultra-small grains (Draine \& Lazarian 1998). These ultra-small
grains are thought to exist mainly in the form of single polycyclic aromatic
hydrocarbon (PAH) molecules. PAH molecules are generally detected through their
narrow line emission features in the MIR. For these emission features to be
observed the PAH molecules must be exposed to a strong source of UV flux. Since
this flux is generally absent in the case of dark clouds, the microwave
emission from the rotation of PAH molecules may be the only way to study the
very small grain population in these objects.

It is also known that radio continuum emission in dark clouds may arise from
ionized gas associated with a stellar outflow. When a luminous star is present
this arises either as the result of a compact {\sc{Hii}} region or an ionized
stellar wind. In the case of very young low luminosity stars radio continuum
emission may be also be detected. In this instance it is generally attributed
to the presence of a partially ionized ($0.02\leq x_{\rm e} \leq 0.35$;
Bacciotti \& Eisl{\"o}ffel 1999) stellar wind (Wright \& Barlow 1975; Panagia
\& Felli 1975), or possibly a neutral wind which has been shock-ionized further
from the central source by impacting on a dense obstacle (Curiel et~al.\ 1989).

In this paper we present follow-up observations of the five AMI Small Array
(SA) spinning dust detections (Paper I) at higher resolution with the AMI Large
Array (LA) over the same frequency range. All co-ordinates in this paper are
J2000.0.

\section{Observations}

AMI comprises two synthesis arrays, one of ten 3.7\,m antennas (SA) and one of
eight 13\,m antennas (LA), both sited at Lord's Bridge, Cambridge (AMI
Consortium: Zwart et~al.\ 2008). The telescope observes in the band
13.5--17.9\,GHz from which eight 0.75\,GHz bandwidth channels are synthesized.
In practice, the two lowest frequency channels (1 \& 2) are not generally used
due to a lower response in this frequency range and interference from
geostationary satellites.

Observations of five Lynds dark nebulae selected from the original AMI SA
sample were made in 2009 February--March using the AMI LA. The co-ordinates of
these fields are listed in Table~\ref{tab:lynds} along with the size of the AMI
LA synthesized beam towards each object and the r.m.s.\ noise measured outside
the primary beam on the CLEANed maps. We note that the AMI LA observation of
L1246 is towards the north--east of this cloud where anomalous emission was
detected by the AMI SA and does not cover the same area as the original SCUBA
observation.

Data reduction was performed using the local software tool \textsc{reduce}, see
Paper I for more details. Flux calibration was performed using short
observations of 3C286 near the beginning and end of each run. We assumed I+Q
flux densities for this source in the AMI LA channels consistent with the
frequency dependent model of Baars et~al.\ (1977), $\simeq 3.3$\,Jy at 16\,GHz.
As Baars et~al.\ measure I and AMI LA measures I+Q, these flux densities
include corrections for the polarisation of the calibrator source derived by
interpolating from VLA measurements. A correction is also made for the changing
intervening air mass over the observation. The phase was calibrated using
interleaved observations of calibrators selected from the Jodrell Bank VLA
Survey (JVAS; Patnaik et~al.\ 1992). After calibration, the phase is generally
stable to $5^{\circ}$ for channels 4--7, and $10^{\circ}$ for channels 3 and 8.
The FWHM of the primary beam of the AMI LA is $\approx 6$\arcmin at 16\,GHz.

\begin{figure*}
\centerline{L675\hspace{8.5cm}L944}
\centerline{\includegraphics[width=8.5cm,clip=,angle=0.]{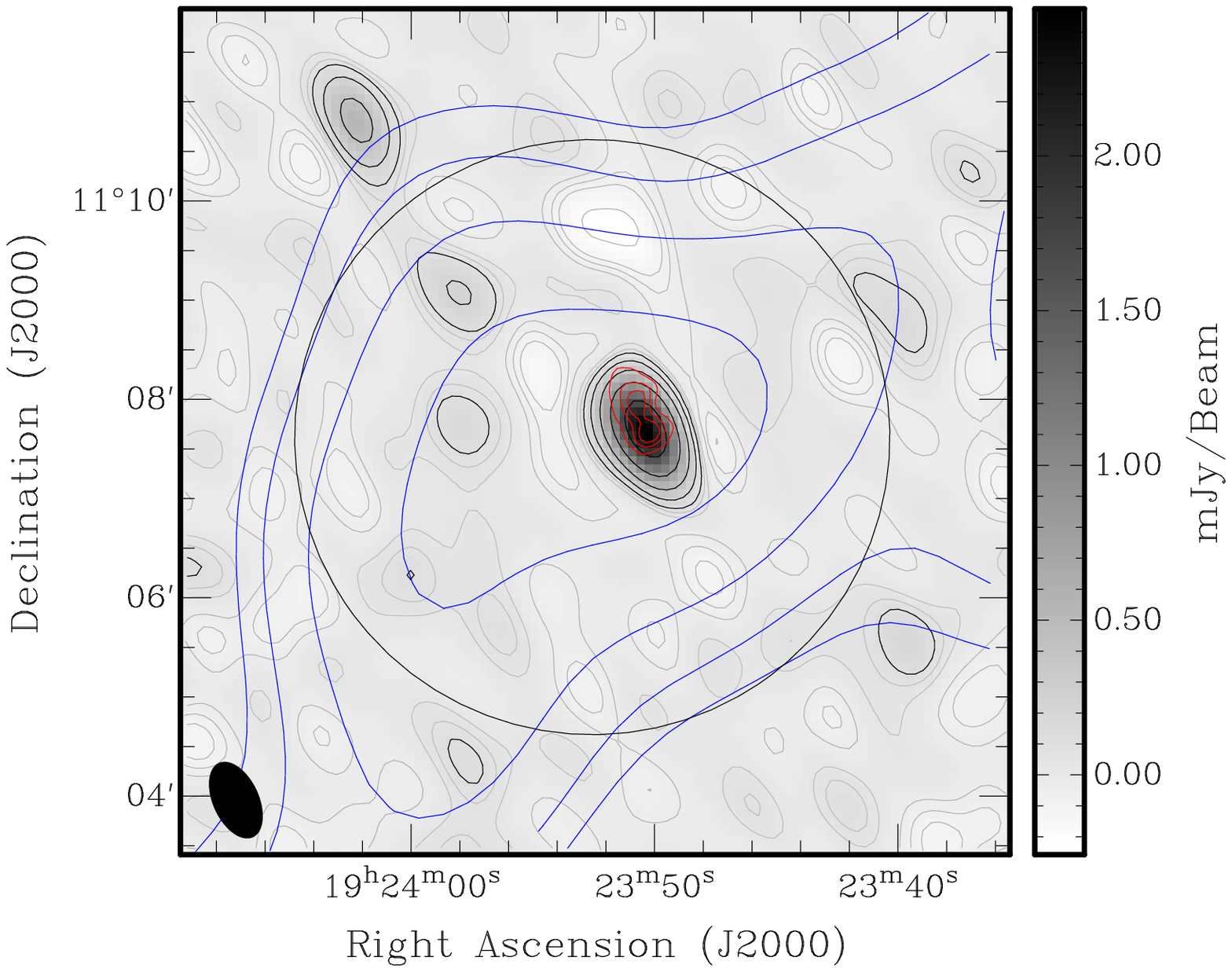}\qquad\includegraphics[width=8.5cm,clip=,angle=0.]{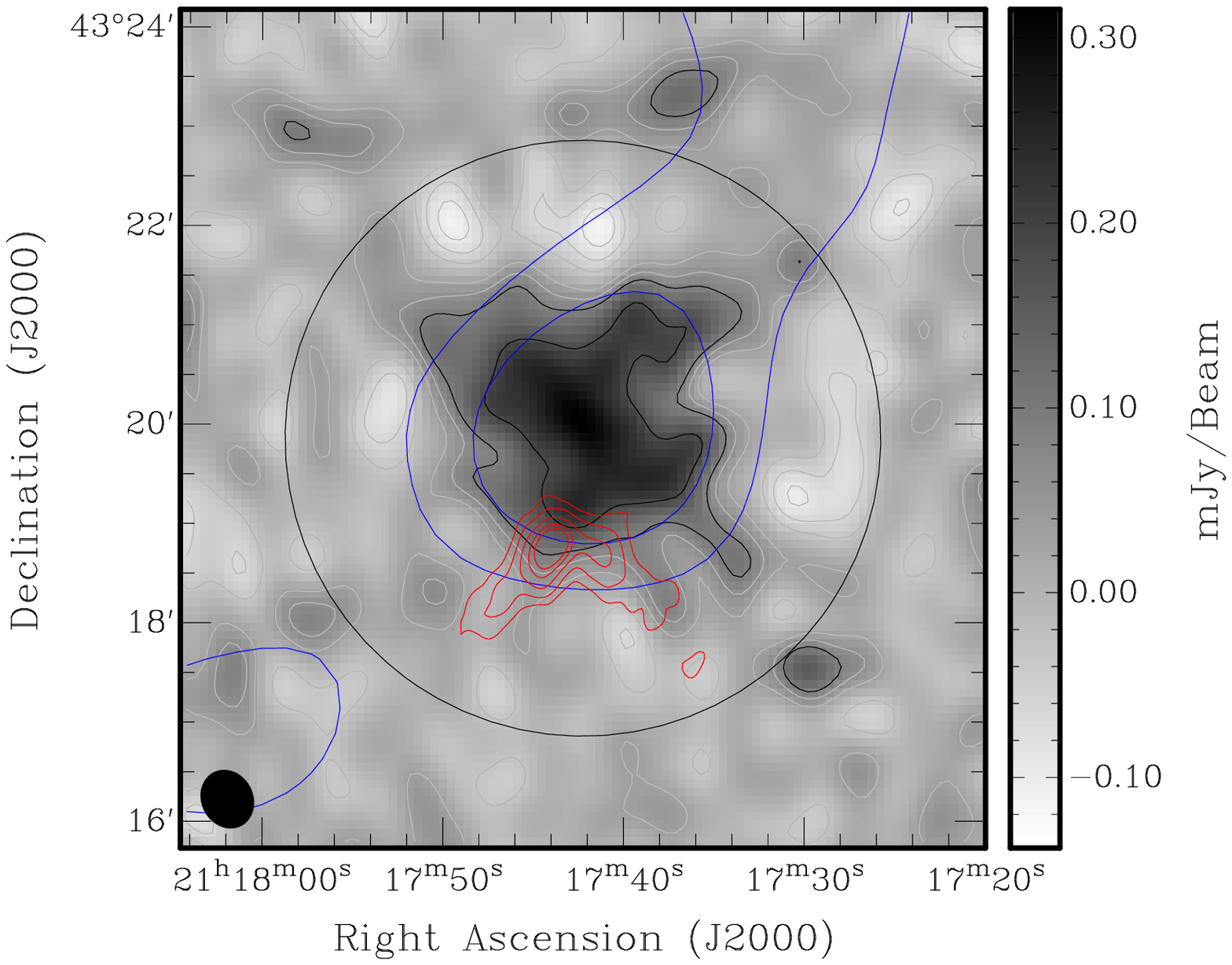}}
\centerline{L1103\hspace{8cm}L1111}
\centerline{\includegraphics[width=8.5cm,clip=,angle=0.]{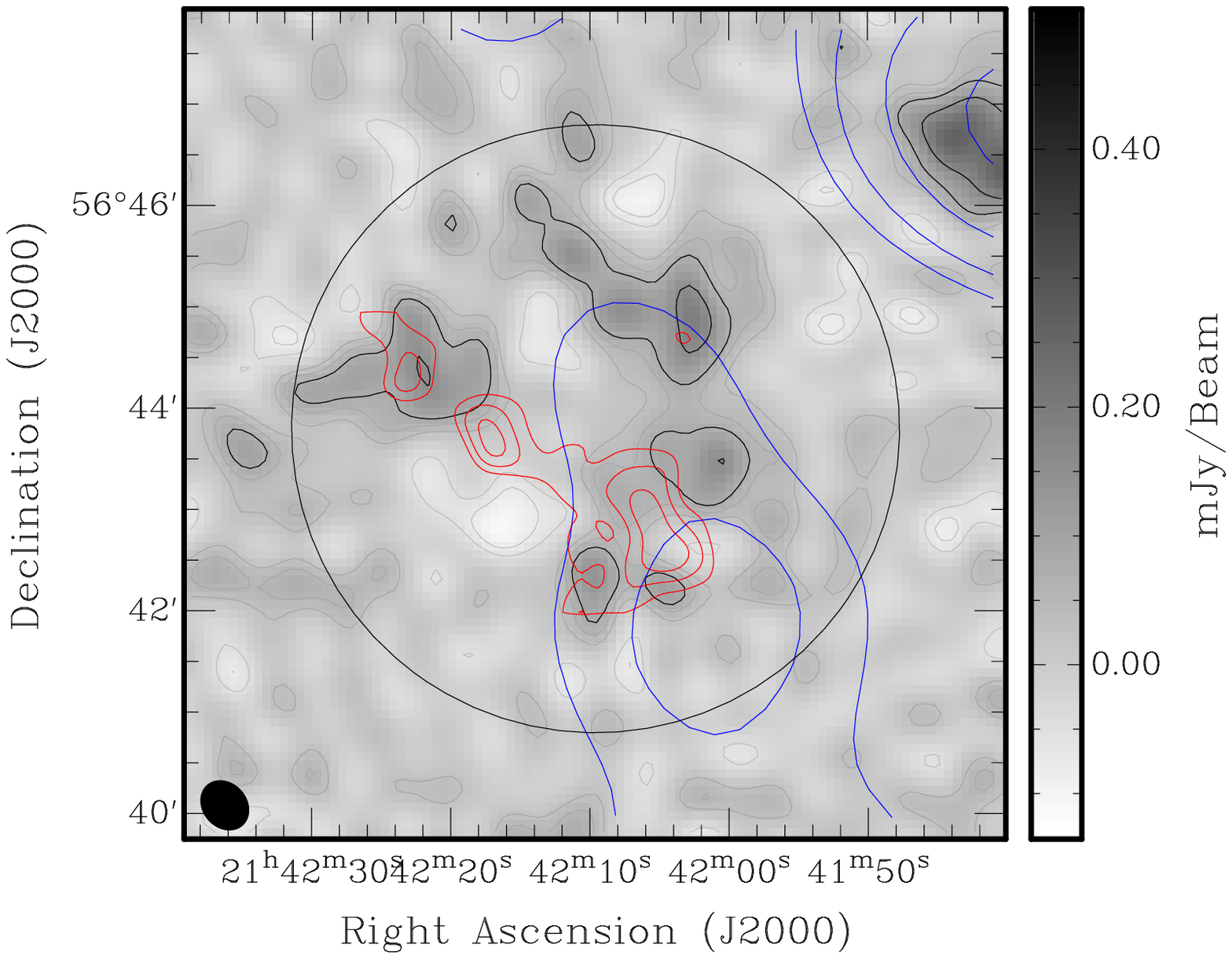}\qquad\includegraphics[width=8.5cm,clip=,angle=0.]{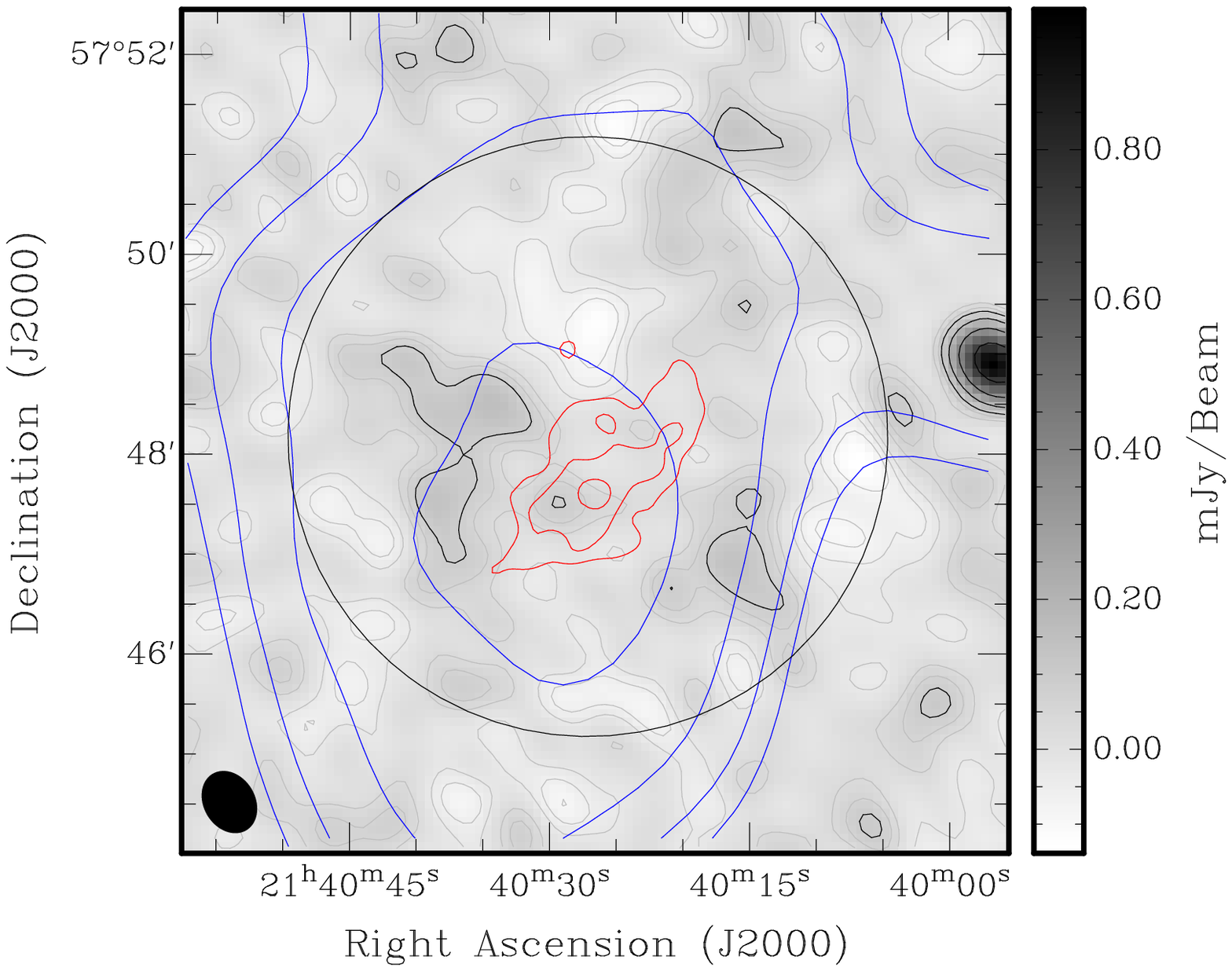}}
\centerline{L1246}
\centerline{\includegraphics[width=8.5cm,clip=,angle=0.]{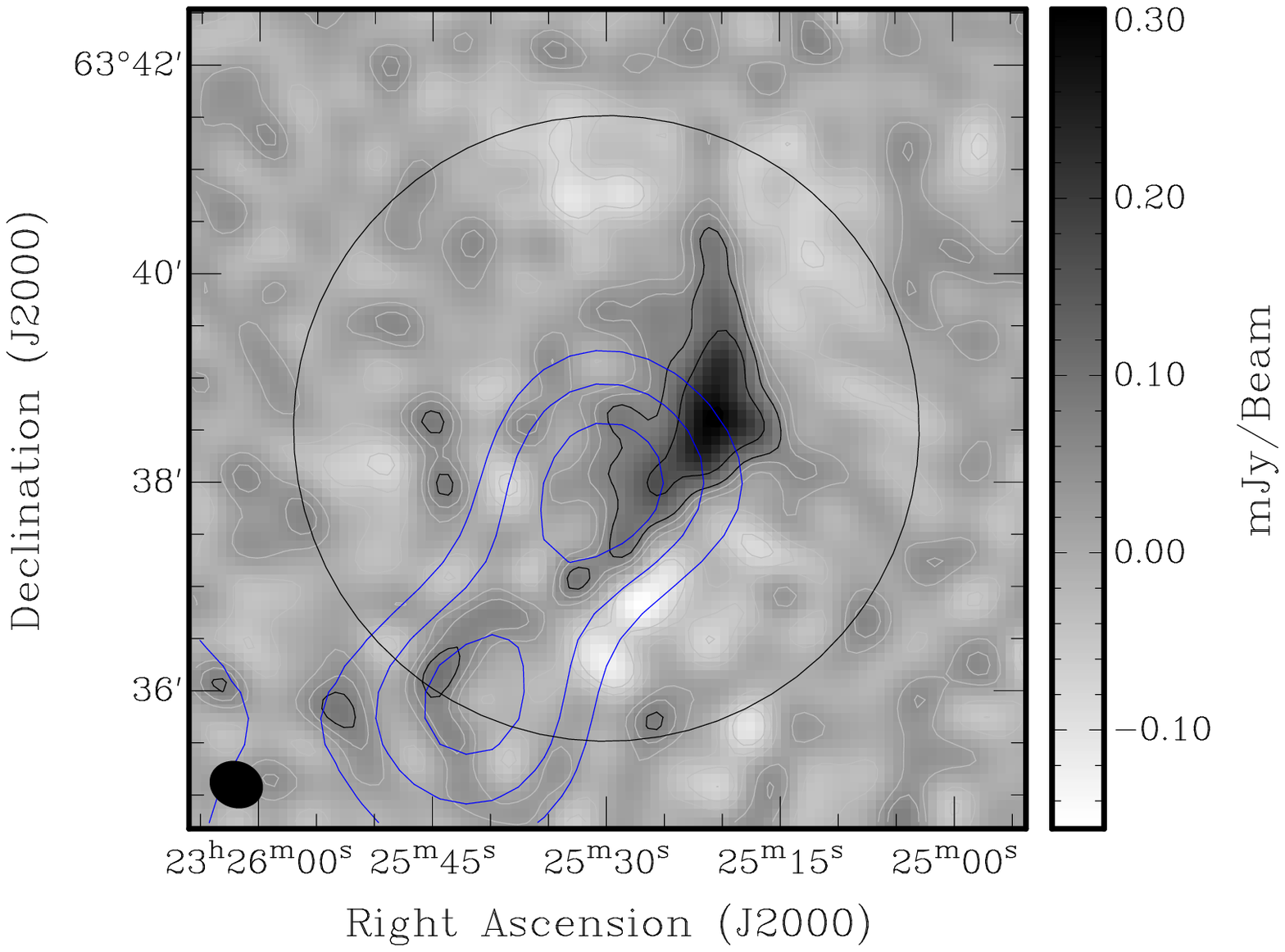}}
\caption{AMI LA combined channel data is shown as greyscale in units of
mJy/beam, grey conrtours at $-6,-3,\pm2,\pm1 \sigma$ and black contours at 3, 6, 12, 24\,$\sigma$ etc. SCUBA 850\,$\mu$m
data is shown as red contours with levels as in Visser et~al.\ (2001; 2002) for
the all clouds except L1246. The AMI LA observation of L1246 does not cover the
region observed by SCUBA. AMI SA data is shown as blue contours with levels as
in Paper I. The AMI LA primary beam FWHM is shown as a circle and the
synthesized beam as a filled ellipse in the bottom left
corner.\label{fig:la_data}}
\end{figure*}

Reduced data were imaged using the AIPS data package. {\sc{clean}}
deconvolution was performed using the task {\sc{imagr}} which applies a
differential primary beam correction to the individual frequency channels to
produce the combined frequency image. Deconvolved maps were made
from both the combined channel set, see Fig.~\ref{fig:la_data}, and for individual channels. The broad
spectral coverage of AMI allows a representation of the spectrum between 14.3
and 17.9\,GHz to be made independently of other telescopes and in what follows
we use the convention: $S\propto \nu^{-\alpha}$, where $S$ is flux density,
$\nu$ is frequency and $\alpha$ is the spectral index. All errors are quoted to
1\,$\sigma$.

\begin{table}
\caption{AMI LA Lynds Dark Nebulae. Column [1] Name of cloud, [2] Right
Ascension, [3] Declination, [4] AMI LA synthesized beam FWHM major axis, [5]
AMI LA synthesized beam FWHN minor axis, and [6] r.m.s.\ noise fluctuations on
the combined channel map. \label{tab:lynds}}
\begin{tabular}{lccccc}
\hline \hline
Name & RA  & Dec  & $\Delta \theta_{\rm{maj}}$ & $\Delta
\theta_{\rm{min}}$ & $\sigma_{\rm{rms}}$\\
     & (J2000) & (J2000) & (arcsec) & (arcsec) & ($\frac{\mu{\rm{Jy}}}{\rm{bm}}$)\\\hline
L675 & 19 23 52.6 & 11 07 39 & 49.9 & 27.4 & 35 \\
L944 & 21 17 40.8 & 43 18 08 & 36.5 & 31.2 & 31 \\
L1103 & 21 42 10.2 & 56 43 44 & 32.0 & 26.3 & 25 \\
L1111 & 21 40 27.1 & 57 48 10 & 39.4 & 30.8 & 29 \\
L1246 & 23 25 30.1 & 63 38 30 & 31.2 & 26.9 & 25 \\\hline
\end{tabular}
\end{table}

\section{Results}

\begin{table*}
\caption{Integrated flux densities in mJy for AMI LA observations of L675, L944
and L1246. Errors are calculated as $\sigma =
\sqrt{(0.05S)^2+\sigma_{\rm{rms}}^2}$, where $\sigma_{\rm{rms}}$ is the r.m.s.\
noise in the individual channel map.\label{tab:flist}}
\footnotesize
\begin{tabular}{lccccccccccccc}\hline\hline
&\multicolumn{12}{c}{Freq. (GHz)}& \\\cline{2-13}
Name&14.3&&15.0&&15.7&&16.4&&17.2&&17.9&&$\alpha$\\
\hline
L675&2.74&$\pm0.34$&2.24&$\pm0.12$&2.26&$\pm0.12$&2.12&$\pm0.11$&2.42&$\pm0.14$&2.22&$\pm0.12$&$+0.10\pm$0.36\\
L944&-&&2.03&$\pm0.11$&2.40&$\pm0.13$&2.89&$\pm0.15$&2.86&$\pm0.15$&-&&$-2.11\pm0.49$\\
L1246&-&&0.62&$\pm0.18$&0.55&$\pm0.03$&0.62&$\pm0.03$&0.58&$\pm0.04$&-&&$-0.40\pm0.82$\\
\hline
\end{tabular}
\normalsize
\end{table*}

\textbf{L675:} The AMI LA observations of L675 show two obvious regions of
compact emission, see Fig.~\ref{fig:la_data}. The first of these, slightly
offset from the pointing centre, is coincident with both the peak of the AMI SA
emission and also the compact emission seen at 850\,$\mu$m by the SCUBA
instrument (Visser et~al.\ 2001; 2002). We denote this source ``A'' ($19^{\rm
h}23^{\rm m}50\fs5$, $+11^{\circ}07'44''$). The second, just outside the LA
primary beam FWHM to the north-east, is coincident with the probable
extragalactic point source identified as ``B'' ($19^{\rm h}24^{\rm m}02\fs4$,
$+11^{\circ}10'54''$) in the original AMI SA observations (Paper I).

\begin{figure}
\centerline{\includegraphics[width=4cm,height=7cm,angle=-90]{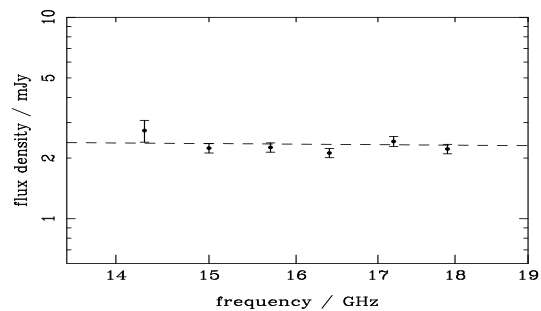}}
\caption{L675 Source A: data points are flux densities from AMI LA channels
3--8. The best-fit spectral index of $\alpha=0.10\pm0.36$ is shown as
a dashed line.\label{fig:l675spec}}
\end{figure}

Source A is completely unresolved by the AMI LA and shows a flat spectrum
across the AMI band, $\alpha_{14.3}^{17.9}=0.10\pm0.36$, consistent with
free--free emission, see Fig.~\ref{fig:l675spec}. This spectral index differs
considerably from that measured by AMI SA. This is because the LA is not
sensitive to the large scale emission seen with the SA. Indeed it seems likely
that the emission seen by the two arrays arises from completely different
sources.

\textbf{L944:} The original AMI SA observations of L944 revealed a compact
region of emission to the north of the cloud, coincident with one side of the
protostellar outflow. AMI LA observations, see Fig.~\ref{fig:la_data}, reveal
this emission arises not from a point-like object but rather from a diffuse
region of emission, the peak of which occurs at $21^{\rm h}17^{\rm m}42\fs6$,
$+43^{\circ}20'08''$. We estimate the flux spectrum by integrating the flux
density from the primary beam corrected channel maps within a two arcminute
radius of the LA pointing centre. This shows a steeply rising spectrum with
$\alpha_{14.3}^{17.9} = -2.1\pm0.5$. This is consistent with that found from
the AMI SA data, however this correspondence is not meaningful as the low
signal to noise in the SA data precludes a precise estimate. The flux density
found towards this region in the LA map is only marginally lower than that
found from the comparatively coarser resolution SA map. This implies that the
emission comes not from one smooth extended region that is partially resolved
out by the LA baselines, but from a collection of smaller fragments or
filaments. These fragments are unresolved by either array, although the
granularity becomes more evident in the higher frequency channels of the LA.
The amount of flux lost in channels 5 to 8 relative to channel 4 is
significantly smaller than would be expected from a Gaussian source of similar
dimensions.

\textbf{L1103 and L1111:} AMI LA observations of L1103 and L1111 do not show
any distinct regions of compact cm-wave emission. The diffuse patches of low
level emission present within the primary beam towards both sources are
indicative of larger scale structures which have been resolved out by the
synthesized beam. We can provide an estimate of the flux density seen towards
these objects with the LA by fitting and removing a tilted plane baselevel at
the primary beam FWHM. From the combined channel data this gives $S_{16}
=5.1\pm0.6$\,mJy and $S_{16} = 2.4\pm0.3$\,mJy for L1103 and L1111,
respectively. These values indicate that the flux loss is considerable:
approximately 45\% and 96\%. The sensitivity of these LA observations is much
greater than those of the SA and it is possible that the patchy emission in
these fields corresponds to enhancements in the extended emission which are
below the detection threshold in the SA data.

\textbf{L1246:} AMI SA observations towards L1246 did not show any excess
emission coincident with the SCUBA identification of the dark cloud, but did
reveal a region of emission $\approx 2'$ to the north--east of the cloud, in a
region not covered by the SCUBA map, which had no counterpart in the lower
frequency observations. AMI LA observations of this NE region show an arc of
emission (peak: $23^{\rm h}25^{\rm m}20\fs4$, $+63^\circ38'40''$), see
Fig.~\ref{fig:la_data}. We assess the spectral behaviour of this object in two
ways. Firstly, we estimate the flux density of the arc itself. We fit and
remove a tilted plane baselevel within an irregular polygon drawn around the
object and integrate the remaining flux. Secondly, we fit a tilted plane
baselevel to a circle at the primary beam FWHM and integrate all the flux above
this baselevel within that radius. Both methods give consistent results, as
might be expected since the primary beam is relatively empty otherwise. From
the first method we find a spectral index $\alpha_{15.0}^{17.9}=-0.40\pm0.87$,
and from the second $\alpha_{15.0}^{17.9}=-0.47\pm0.82$.

\begin{figure}
\centerline{\includegraphics[width=7cm,clip=,angle=0]{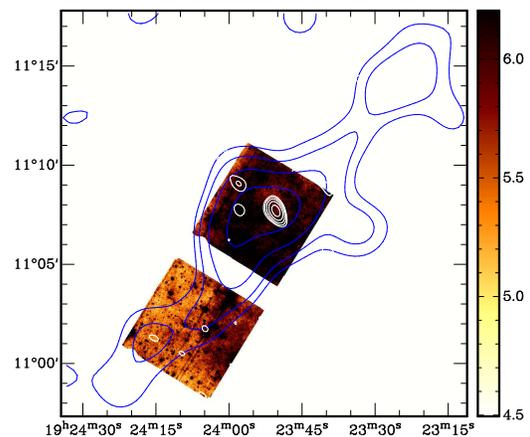}}
\caption{L675: AMI LA combined channel data is shown as white contours at 3, 6,
12\,$\sigma$ etc. \emph{Spitzer} Band 4 data is shown as greyscale in MJy/sr,
and is saturated at both ends to emphasise the structure present. AMI SA data
is shown as blue contours as in Fig.~\ref{fig:la_data}.\label{fig:l675spit}}
\end{figure}

\begin{figure}
\centerline{\includegraphics[width=8cm,clip=,angle=0]{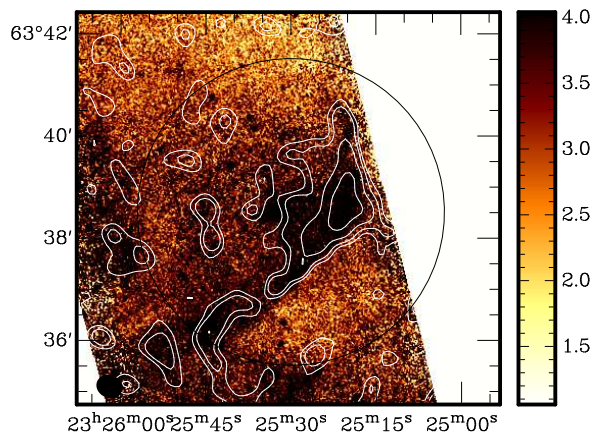}}
\caption{L1246: AMI LA combined channel data is shown as white contours at 1,
2, 4, 8\,$\sigma$ etc. \emph{Spitzer} Band 4 data is shown as greyscale in
MJy/sr, saturated at both ends of the scale to emphasise the structure present.
The AMI LA primary beam is shown as a circle and the synthesized beam as a
filled ellipse in the bottom left corner. \label{fig:l1246spit}}
\end{figure}

\section{Discussion and Conclusions}

L675 and L1246 have archival \emph{Spitzer} IRAC data, which shows in both
cases a significant amount of emission in Band 4 (6.4--9.4\,$\mu$m) and very
little in the other three (3.2--3.9, 4.0--5.0 and 4.9--6.4\,$\mu$m,
respectively). In the case of L675 this emission is present on a very large
scale, see Fig.~\ref{fig:l675spit}. The emission seen at 16\,GHz with the AMI
SA appears on a similar scale, however the small field of view of the
\emph{Spitzer} data precludes a more detailed comparison. L1246 shows an arc of
emission at 16\,GHz which is also evident in \emph{Spitzer} IRAC Band 4, see
Fig.~\ref{fig:l1246spit}. This emission is again not present in Bands 1--3. In
Band 4 it is present as an arc, coincident with that seen at 16\,GHz in the AMI
LA data.

\emph{Spitzer} Band 4 contains two of the PAH emission lines, including the
strongest (7.7\,$\mu$m). Of the three other \emph{Spitzer} bands only Band 1
contains an emission line (3.3\,$\mu$m) and for ionized PAHs this line is
expected to be significantly weaker. It is probable therefore that the MIR
correlated cm-wave data seen in the AMI maps is a consequence of spinning dust
emission from a population of ionized PAH molecules. Neutral PAH molecules do
not in general possess a permanent dipole moment and are therefore not expected to have
rotational emission (Tielens 2008). This emission, the mechanism of which is
described in detail by Draine \& Lazarian (1998), arises from the intrinsic
dipole moments of small dust grains, most likely to be PAH molecules, which
emit power when they rotate. This rotation has a variety of contributing
factors, the relative importance of which varies with grain environment.
However, in the majority of cases excitation through collision with ions is
predominant.

In the case of L675A, we must consider the possibility that we are observing a
coincidental extragalactic radio source. Using the extended 9C survey 15\,GHz
source counts (Waldram et~al.\ 2009), where $n(S) = 51
(S/{\rm{Jy}})^{-2.15}$\,Jy$^{-1}$\,sr$^{-1}$, the probability that a source
with flux density greater than 2\,mJy lies within the FWHM of the AMI LA
primary beam is $0.12$, and only 0.01 within the SCUBA field. It is likely
therefore that the radio source L675A is associated with the SCUBA core.

A further question is whether the cm-wave emission might be explained by
thermal (Planckian) dust emission. A single greybody spectrum with a dust
temperature, $T_{\rm{d}} \approx 27$\,K, might be used to explain the LA flux
density, however it would require a $\beta$ of 0.6. Such a value would be
unusual even for objects known to possess flattened dust tails, such as
protoplanetary disks. This simple fit also neglects the flux lost by the AMI LA
baseline distribution. SA observations have already shown this source to
possess a significant amount of extended emission which would make this
scenario even more unlikely.

The presence of a neutral or partially ionized wind from an outflow source that
has been shocked through encountering a dense obstacle (Torrelles et~al.\ 1985;
Rodr{\'i}guez et~al.\ 1986) is used to understand the spectral indices seen
towards exciting sources in the radio regime (Curiel et~al.\ 1990; Cabrit \&
Bertout 1992). This model allows a spectral index range of $0.1$ (optically
thin) to -2 (optically thick), which explains results which deviate from the
value of $\alpha = -0.6$ required by a spherically symmetric ionized wind
(Wright \& Barlow 1975; Panagia \& Felli 1975). Using this model as described
in Curiel et~al.\ (1989; 1990) the radio emission is expected to be optically
thin ($\tau=0.1$), consistent with the spectral index seen across the AMI band.
Assuming a distance of 300\,pc and a stellar wind with a wind speed of
200\,km\,s$^{-1}$, we can calculate that the AMI flux densities towards L675A
are consistent with a mass loss of $3.5\times 10^{-7}$\,M$_{\odot}$\,yr$^{-1}$.
A mass loss such as this implies a mechanical luminosity from the wind of
$L_{\rm{mech}} \approx 1.1$\,L$_{\odot}$, comparable to the values found by
Curiel et~al for L1448.

The nature of the emission seen towards L944 with the AMI LA is uncertain. The
spectral index of this emission is consistent with spinning dust emission or
alternatively the optically thick component of free--free spectrum. Such a
free--free spectrum might be exhibited at 16\,GHz by ultra-compact {\sc Hii}
regions. However a turn-over frequency above 16\,GHz would have an extremely
high mass and should therefore be obvious in sub-mm observations. This needs to
be confirmed by either higher radio frequency measurements in order to measure
the optically thin region of the spectrum and the turn-over, or sub-mm
measurements to place constraints on the mass of such a region.

In conclusion, we have used the AMI LA to observe a sample of five Lynds Dark
Nebulae selected as candidates for spinning dust emission from the AMI SA
sample of Lynds Dark Nebulae (Paper I). Towards two of these clouds (L1103 and
L1111) we detect only patchy diffuse emission characteristic of the presence of
a larger structure which has been mostly resolved out.

Towards L675 we have observed flat spectrum compact cm-wave emission coincident
with the SCUBA 850\,$\mu$m emission from the same region. These characteristics
suggest that this source is associated with a stellar wind from a deeply
embedded young protostar.

We detect extended cm-wave emission to the North of the L944 SMM-1 protostar
which displays spectral behaviour consistent with either spinning dust, or
alternatively a collection of ultracompact {\sc Hii} regions.

L1246 shows an arc of cm-wave emission which is coincident with emission seen
in \emph{Spitzer} Band 4. We suggest that this is an example of emission from a
population of PAH molecules, seen in emission lines in the \emph{Spitzer} data,
and emission as a consequence of rapid rotation of the molecules in the cm-wave
data.

\section{ACKNOWLEDGEMENTS}

We thank the staff of the Lord's Bridge Observatory for their invaluable
assistance in the commissioning and operation of AMI. AMI is supported
by Cambridge University and the STFC. NH-W, CR-G, 
TWS, TMOF, MO and MLD acknowledge the support of PPARC/STFC studentships. This
work is based in part on archival data obtained with the Spitzer Space
Telescope, which is operated by the Jet Propulsion Laboratory, California
Institute of Technology under a contract with NASA. Support for this work was
provided by an award issued by JPL/Caltech.

\bsp
\label{lastpage}

\end{document}